\documentclass[proceedings, preprint]{rmaa}

\usepackage{paralist}

\usepackage{psfrag,color}

\SetYear{2008}
\SetConfTitle{Magnetic Fields in the Universe II}

\title{Magnetic Effects in Global Star Formation} 

\author{Mordecai-Mark Mac Low\altaffilmark{1}}

\altaffiltext{1}{Department of Astrophysics, American Museum of
  Natural History, New York, NY, 10024-5192, USA (mordecai@amnh.org).}

\shortauthor{Mac Low}
\shorttitle{Magnetic Effects in Global Star Formation}

\listofauthors{M.-M. Mac Low}
\indexauthor{Mac Low, M.-M.}

\abstract{I review the effects of magnetic fields on star formation in
galaxies.  This includes the effects of the magnetorotational
instability (MRI) at galactic scales, magneto-Jeans and swing
instabilities, Parker instabilities, and the effects of magnetic
fields on the evolution of supernova-driven turbulence. I argue
that currently turbulent support by the MRI appears likely to be the
most important of these processes to regulating star formation.
}

\addkeyword{magnetic fields}
\addkeyword{stars: formation}
\addkeyword{ISM: magnetic fields}
\addkeyword{galaxies: magnetic fields}

\begin{document}
% Typeset article header
\maketitle

\section{Global Star Formation}
The question of what regulates star formation in galaxies remains
unsolved, with at least four main scenarios, and a number of
variations on some of them.  To understand the role of magnetic
fields, we must first briefly review these ideas.

The first is that global star formation is primarily controlled by
gravitational instability of the available gas in the disk.  Whether
this instability is global, controlled by the combined potential of
the gas and stars \citet{rafikov01}, or local, controlled by the
behavior of individual GMCs, is still argued.  The basic idea of
global instability was described by \citet{elmegreen02}, and supported
by numerical models \citep{kravtsov03,lmk05}. Observational evidence
for a direct correlation between star formation rate and gravitational
instability in the LMC is given by \citet{yang07}.  Local instability
in molecular clouds has been argued to be the rate limiting step by
\citet{krumholz05}.

%% A second idea that remains popular is sequential, triggered, star
%% formation, as originally proposed by \citet{gerola77} and
%% \citet{elmegreen-lada77}.  Although second and even third generation
%% star formation around the edges of star forming regions now appears to
%% be a well-established phenomenon, both observations \citep{mizuno07}
%% and theory \citep{jm06} seem to suggest that it is typically a 10\%
%% effect, and unimportant for the large-scale picture.

A second idea recently elucidated by \citet{shu07} is that
magnetic regulation dominates star formation.  The results of
\citet{kim06} and \citet{shetty06} establish that the magnetic
field can control the morphology of star formation in spiral
galaxies. The suggestion is then made by \citet{shu07} that the rate
limiting step in star formation is the accumulation of sufficient mass
to form a supercritical cloud.  

Star formation controlled by a threshold column density has been
advocated by \citet{schaye04}.  He argues that gas above the threshold
will be able to cool quickly by forming molecules, which will drive
gravitational instability. However, this would predict uniformly high
molecular fractions at the critical column density, contrary to the
observations of \citet{mk01}.

Finally, self-regulated star formation remains on the table
\citep[e.g.][]{silk97}. In this scenario, the supernovas produced by
star forming regions determine the level of turbulence in the
surrounding gas \citep[e.g.,][]{slyz05}, which determines its level of
gravitational instability and thus how many stars it can form.  Less
star formation lowers the level of turbulence, which in turn increases
the star formation rate.  However, this cannot explain the lack of
star formation far from star forming regions and massive stars, e.g.\ in the
outer disks of normal galaxies, or in low surface brightness galaxies.

\section{Gravitational Instability}
Global disk instability models postulate that star formation happens
wherever gravitational instability \citep{gammie92,rafikov01} of the
combined collisionless stars \citep{toomre64} and collisional gas
\citep{goldreich65} in the disk sets in during the collapse of disk
galaxies. For a gas disk, the criterion for instability is
\begin{equation}
Q_g\equiv\frac{\kappa c_g}{\pi G \Sigma_g} < 1,
\label{Qg}
\end{equation}
where $\kappa$ is the epicyclic frequency, $c_g$ the speed of sound,
and $\Sigma_g$ the surface density of the gas disk. For a
collisionless stellar disk, the equivalent criterion is
\begin{equation}
Q_s\equiv\frac{\kappa \sigma_s}{\pi G \Sigma_s} < 1.07,
\label{Qs}
\end{equation}
where $\sigma_s$ is the stellar velocity dispersion in the radial
direction, and $\Sigma_s$ is the surface density of the stellar
disk. [Note that following \citet{rafikov01} we use a factor of $\pi$
in the definition of $Q_s$ rather than 3.36, shifting the instability
criterion to slightly higher value.]  Define the dimensionless
quantities $q = k\sigma_s / \kappa$ and $R = c_g/\sigma_s$. Then the
instability criterion for the combined disk of gas and stars is given
by 
\begin{eqnarray}
\frac{2}{Q_s}
\frac{1}{q}\left[1-e^{-q^2} I_0(q^2)\right]
+\frac{2}{Q_g}R\frac{q}{1+q^2 R^2}>1,
\label{dimsg}
\end{eqnarray}
where $I_0$ is the Bessel function of order 0.
The central determining factor in global disk instability
models is then the magnitude of the turbulent support of gas.

\begin{figure*}
\includegraphics[width=\textwidth]{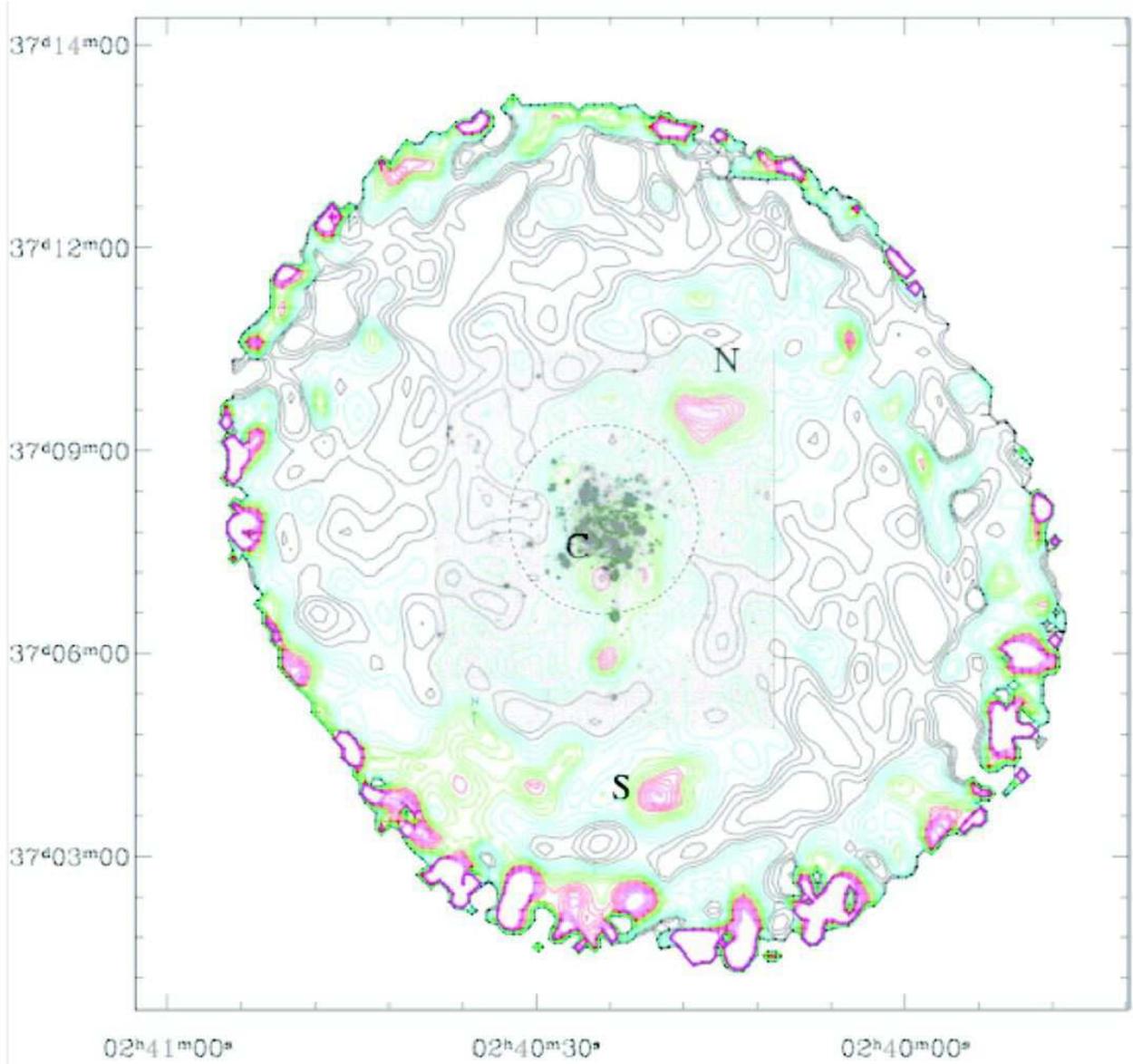}
\caption{\label{petric} Distribution of velocity dispersion of
  \ion{H}{i} in the galaxy NGC~1058, overlaid on the $H\alpha$
  emission from \citet{ferguson98} in greyscale. This demonstrates the
  surprising uniformity and lack of correlation with star formation of
  the velocity dispersion. The regions of highest dispersion are
  labeled N, C, and S. The $x$ and $y$ axis are the RA and Dec in
  B1950 coordinates.  The contours are in km~s$^{-1}$ and start in steps of
  0.5 km~s$^{-1}$. Black is used for dispersions between 5.5 to 7, cyan for
  7.5 to 9, green for 9.5 to 11, red for 11.5 to 13, and magenta for
  13.5 to 15 km~s$^{-1}$. From \citet{petric07}.
%%  These show variations in the velocity dispersion
%%   from 5--15~km~s$^{-1}$ uniformly distributed across the gas disk of
%%   the galaxy, with absolutely no evident correlation with regions of
%%   star formation. The highest velocity dispersion regions are far from
%%   the star-forming central region of the galaxy, which lies in a
%%   perfectly unremarkable region of intermediate dispersion.
}\end{figure*}

This support cannot come solely from star formation, however.  The
21~cm observations of \citet{petric07} represent the
highest-resolution observation of velocity dispersion across a
galactic disk, in this case that of NGC~1058.
(see Fig.~\ref{petric})

This suggests to me that the velocity dispersion of the gas must be
determined by physics more or less independent of star formation. Star
formation might then be simply the response of the turbulent gas disk
to gravity, depending on the amount of gas available from smooth or
lumpy accretion. 

What is the physical mechanism driving the turbulence?
\citet{ml99,ml03} estimates the dissipation rate for a supersonic
turbulent flow to be 
\begin{eqnarray} \label{edot}
\dot{e} &= & -\frac12 \frac{\rho v_{\mbox{rms}}^3}{\lambda_D} \\
        & \simeq & 
  -(3 \times 10^{-27} \mbox{ erg cm}^{-3} \mbox{ s}^{-1}) \times
        \nonumber \\
  & \times &
   \left(\frac{n}{1 \mbox{ cm}^{-3}}\right) 
   \left(\frac{v_{\mbox{rms}}}{10 \mbox{ km s}^{-1}}\right)^3
   \left(\frac{\lambda_D}{100 \mbox{ pc}}\right)^{-1}, \nonumber
\end{eqnarray}
where $\rho$ is the average density, $v_{\mbox{rms}}$ the rms
velocity, and $\lambda_D$ the effective driving scale of the
turbulence.  Note in particular the cubic dependence of the required
energy input on the rms velocity.  A factor of two reduction in the
velocity will lead to an order of magnitude lower energy requirement.

\citet{sellwood99} suggested that the magnetorotational instability (MRI)
\citep{balbus91,balbus98} can drive significant turbulence in a
differentially rotating galactic disk.  They estimated the energy
input by noting that the energy input from the MRI 
\begin{equation} \dot{e} = T_{r\phi} \Omega, \end{equation}
where $T_{r\phi}$ is the Maxwell stress tensor, and $\Omega$ the
angular velocity.  Numerical models by \citet{hawley95} suggest that
$T_{r\phi} = 0.6 B^2 / 8\pi$, so 
\begin{eqnarray}
\dot{e} & = & (1.2 \times 10^{-28}  \mbox{ erg cm}^{-3} \mbox{
  s}^{-1}) \times \\
  & \times & \left(\frac{B}{6 \mbox{ $\mu$G}}\right)^2
\left(\frac{\Omega}{[220 \mbox{ Myr}]^{-1}}\right). \nonumber
\end{eqnarray}
We can solve equation~(\ref{edot}) for the rms velocity that could be
powered by this level of energy input, getting $v_{\mbox{rms}} \sim
4$~km~s$^{-1}$, just a little under the typical values observed by
\citet{petric07}. 

\begin{figure*}
%Piontek & Ostriker figure
\begin{center}
\includegraphics[width=0.22\textwidth]{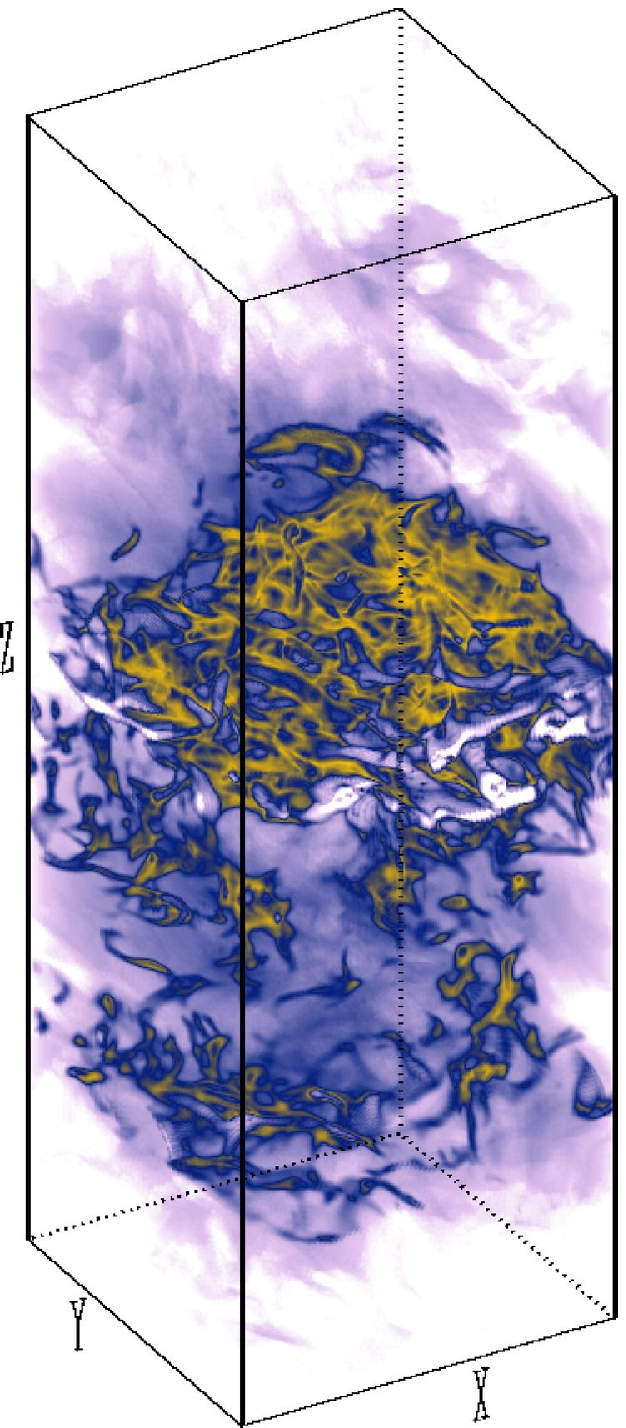}
\includegraphics[width=0.7\textwidth]{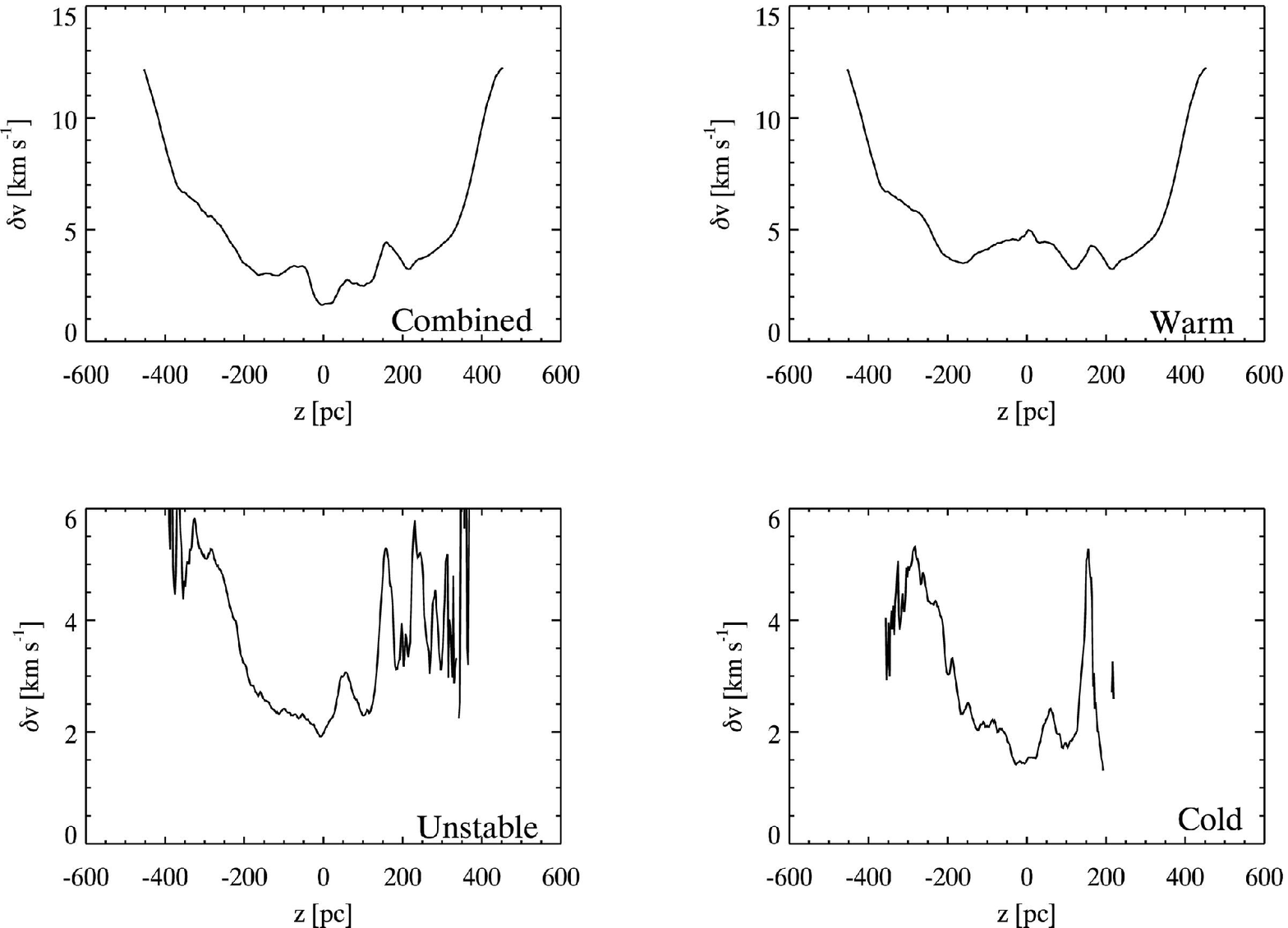}
\end{center}
\caption{\label{piontek}
{\em Left:} Density in a stratified, two-phase, magnetorotationally
unstable disk, computed in a shearing-sheet approximation.
{\em Right:} Vertical profile of mass-weighted velocity dispersion
$\delta v$ for this model for gas within and between the stable
thermal phases, as well as for the total gas.
From \citet{piontek07}.
}
\end{figure*}

Several groups have now simulated the MRI in galactic
disks. \citet{DER04} did three-dimensional global models with an
adiabatic equation of state, and found velocity dispersions varying
with height from 1.5~km~s$^{-1}$ in the midplane to 3~km~s$^{-1}$ at
500~pc and above. \citet{kim03,piontek04,piontek05} and
\citet{piontek07} performed a series of models using a shearing box in
which they moved from an adiabatic equation of state to the
introduction of thermal instability to the inclusion of both thermal
instability and vertical stratification.  They concluded that the
partition of the medium into warm and cold phases was vital to
maintaining significant velocity dispersions, with the velocity
dispersion in the warm gas ranging from 5~km~s$^{-1}$ at the midplane
to over 10~km~s$^{-1}$ at 500~pc, as shown in Figure~\ref{piontek}.

The MRI seems to provide a floor to the possible velocity dispersion
in galactic disks.  This may explain the success of models that follow
gravitational instability in disks of stars and isothermal gas
(embedded in dark matter halos), which can reproduce the Schmidt Law
\citep{lmk05,lmk06}, and show a tight exponential correlation between
the minimum value of the gravitational instability parameter $Q_{sg}$
\citep{rafikov01} and the star formation timescale \citep{lmk05b}.
\begin{figure*}
%LMK figure

\begin{center}
\includegraphics[width=0.5\textwidth]{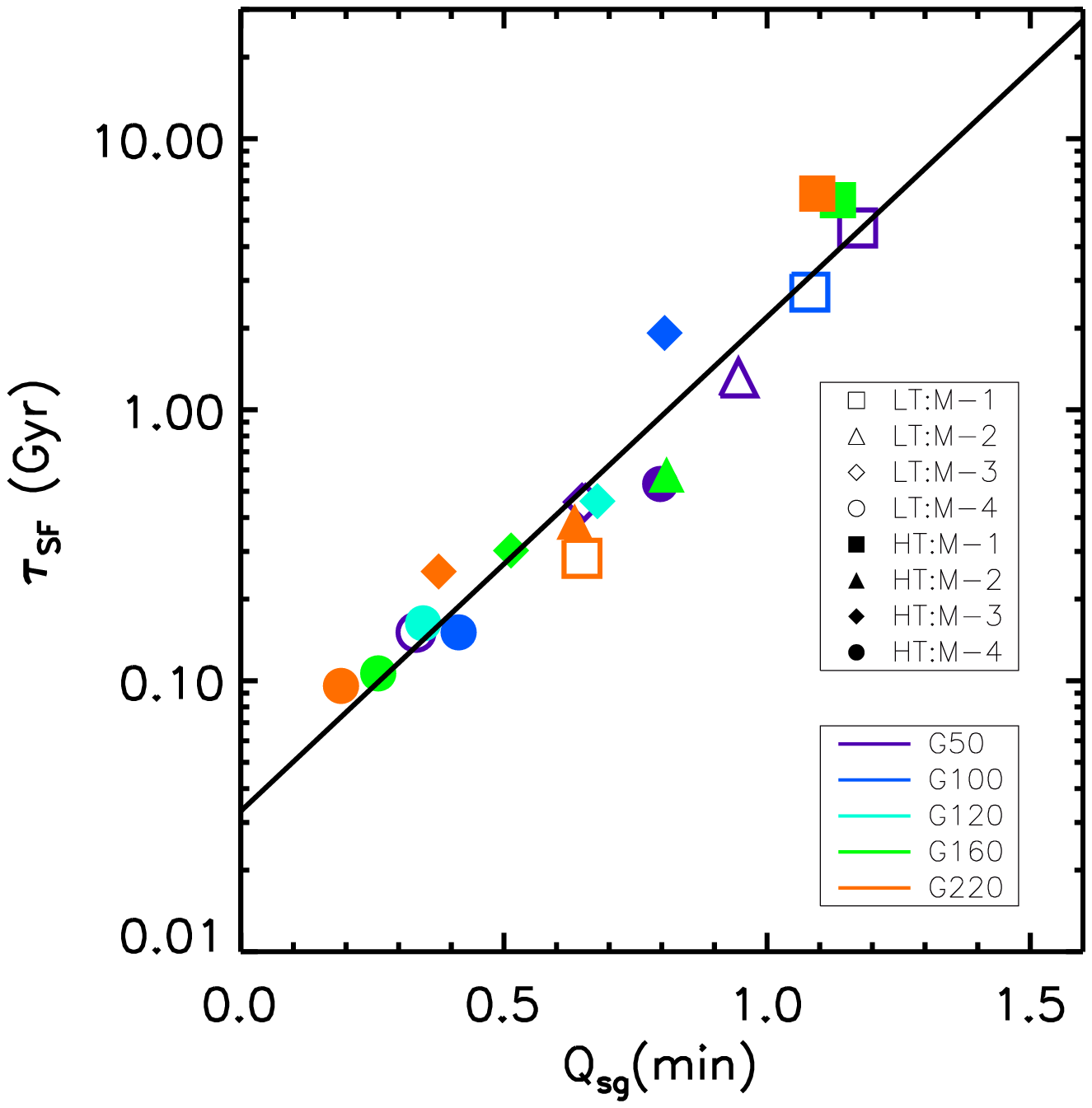}
\includegraphics[width=0.4\textwidth]{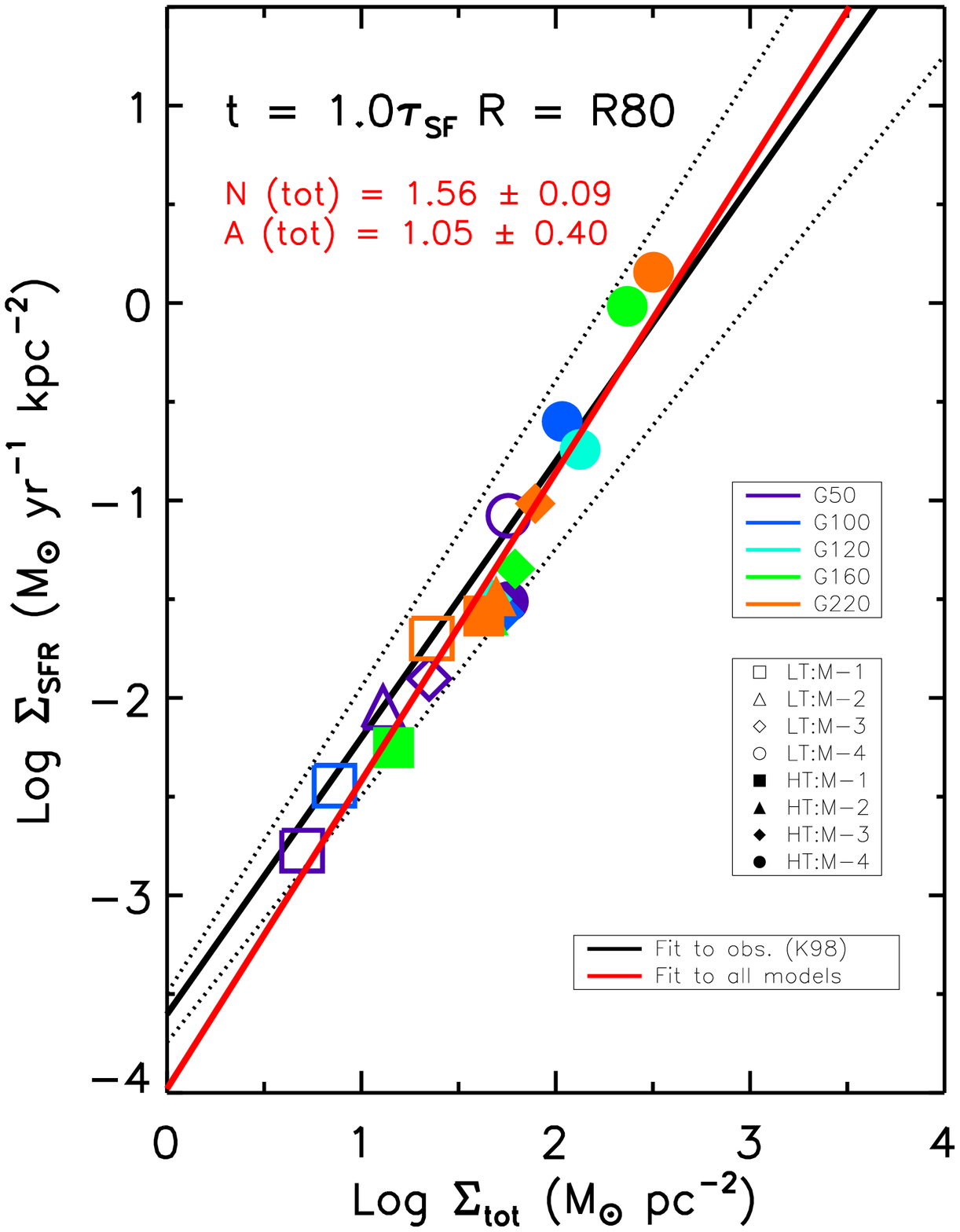}
\end{center}
\caption{\label{lmk} {\em Left:} Star formation timescale $\tau_{\rm
SF}$ correlates exponentially with the initial disk instability
$Q_{sg}$ for both low temperature ({\em open}) and high temperature
({\em filled}) models. The solid line is a least-squares fit. From
\citet{lmk05b}.  {\em Right:} A comparison of the global Schmidt laws
between simulations from \citet{lmk06} and observations. The red line
is the least-square fit to the total gas surface density of the
simulated models, the black solid line is the best fit of
observations from \citet{kennicutt98}, while the black dotted lines
indicate the observational uncertainty. The color of the symbol
indicates the rotational velocity for each model; labels from M-1 to
M-4 are sub-models with increasing gas fraction; and open and filled
symbols represent low and high temperature models, respectively.

} 
\end{figure*}

\section{Magnetic Regulation}
Aside from driving the MRI, magnetic fields can also directly affect
the location and perhaps the rate of star formation in galaxies.
\citet{dobbs07} computed models without self-gravity of multiphase gas
passing through a spiral potential with and without magnetic fields.
They found that, without magnetic fields, clear spurs formed along
their spiral arms, while with fields having magnetic pressure
exceeding the thermal pressure distinct spurs were no longer formed,
though some interarm structure remained. 

\begin{figure*}
\begin{center}
\includegraphics[width=0.8\textwidth]{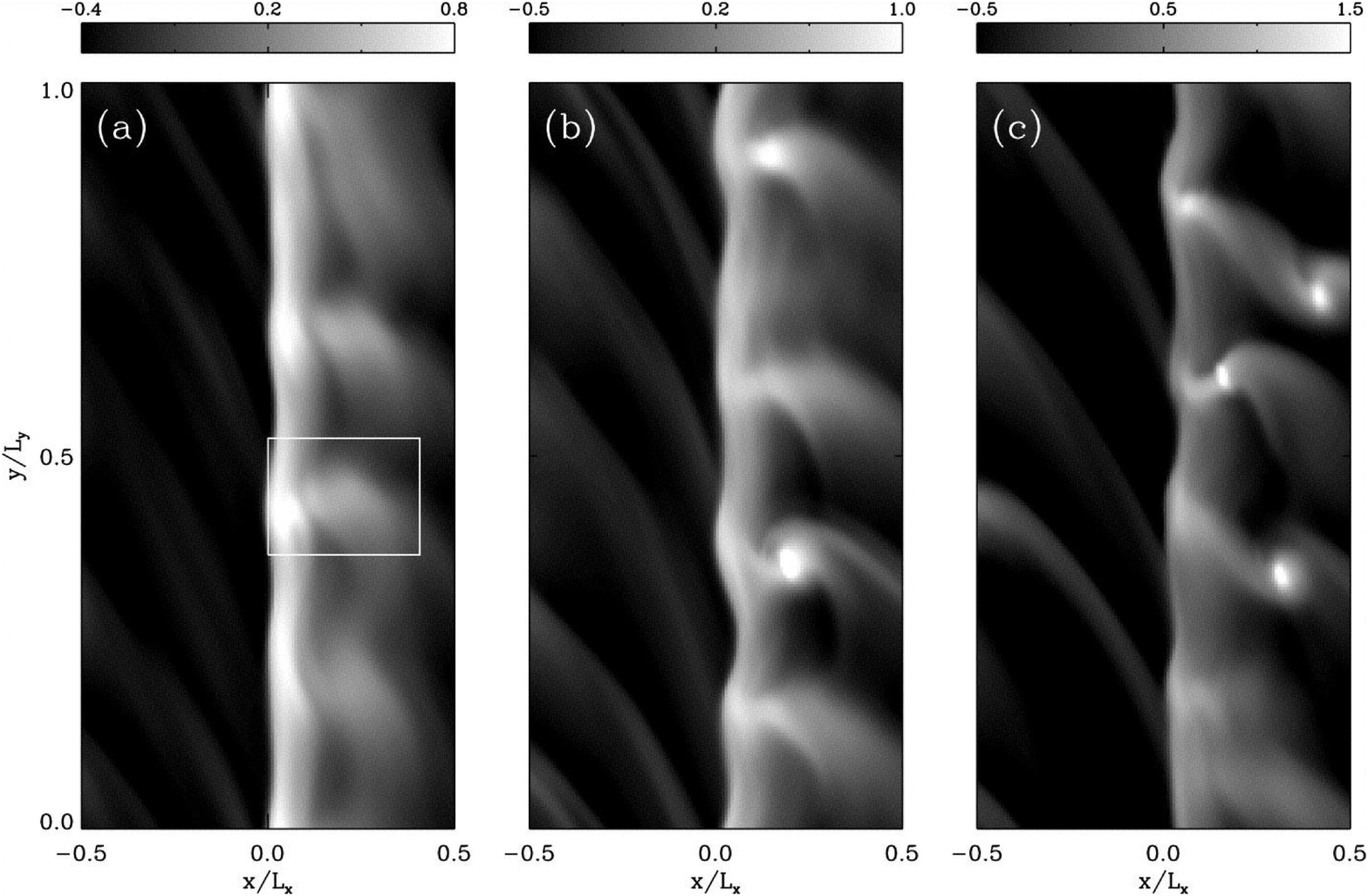} \end{center}
\caption{\label{kim} Surface density ($\log \Sigma / \Sigma_0$ in grey
scale) at times of 5.6, 6, and 6.3 orbital times in the magnetized, 3D
shearing sheet model of \citet{kim06}.  The scale length for typical
parameters is $L_x = L_y/2 = 3.14$~kpc.  }
\end{figure*}

\citet{kim06} performed shearing box models of self-gravitating,
magnetized gas interacting with the fixed stellar potential of a
single spiral arm, an approximation appropriate to a disk dominated by
stellar mass such as that of the Milky Way.  They demonstrated spur
formation and gravitational collapse as natural consequences of the
interaction of fields with self-gravity, as shown in
Figure~\ref{kim}. \citet{shetty06} followed up on these models with
global two-dimensional models of self-gravitating, magnetized gas in a
disk with (and without) an imposed stellar spiral potential,
demonstrating the formation of spurs and self-gravitating objects that
look likely to form giant molecular clouds and ultimately stellar
clusters.

In a magnetized disk, collapse can only occur in regions where the gas
accumulation length along field lines is shorter than the generalized
Toomre instability length in that region.  Cloud formation by such
collapse will then lead to the surrounding gas having reduced
mass-to-flux ratio, and thus resisting collapse more strongly.  This
feedback mechanism can be thought of as magnetic regulation of star
formation \citep{shu07}.  

In spiral galaxies, gas accumulation occurs in spiral arms, so the
work of \citet{kim06} and \citet{shetty06} can be used to derive a
quantitative estimate of star formation in a magnetically regulated
galaxy.  \citet{shu07} demonstrate that they can derive a Schmidt Law
dependence of the star formation rate on gas surface density from
considering spiral arms as example accumulation mechanisms.  However,
this is a special case for the general question of star formation in
galaxies.  It remains to be seen whether this mechanism can explain
the behavior of irregular galaxies without prominent spiral arms,
starburst galaxies, or high-redshift, gas-rich objects.

\section{Galaxy Formation}

Recently the effect of magnetic fields during the galaxy formation era
has been directly studied in a preliminary adaptive mesh refinement
simulation by \citet{wang08}. They studied the collapse of an isolated
halo of total mass $10^{10}$~M$_{\odot}$ initially having an NFW
profile \citep{navarro96}. In this first model, further collapse was
prevented when the Jeans scale approached the grid scale $\Delta x =
26$~pc, so no mass was transferred to a collisionless population of
stars during the evolution of the galaxy.  As a result, gravitational
stirring of the disk is efficient, maintaining velocity dispersions of
10~km~s$^{-1}$ or more, but not from an astrophysically relevant
mechanism.  Magnetic field was assumed to initially thread the halo
with strength of a nanogauss, under the assumption of field production
by previous generations of star formation and active galactic nuclei
\citep[e.g.][]{rees06}.

The initial evolution of the galaxy shows rapidly increasing amounts
of gravitationally unstable gas that ought to produce large numbers of
stars in the first several hundred Myr.  During this period, the field
is amplified in the disk, reaching a few percent of equipartition
after about 500~Myr. The field only begins to play a role in
supporting the gas after that point, however, while star formation
proceeds with little difference between pure hydrodynamic and
magnetized models for the entire previous period.  The initial
conclusion appears to be that even with relatively large initial
fields, magnetic effects can be neglected during the initial period of
galaxy formation, until a large-scale field has been generated.

\section{Field Generation}

This brings us to the question of how such large-scale magnetic fields
are formed. Classical Alpha-Omega dynamos require many Gyr to generate
observed fields, although somewhat shorter times may be given by a
cosmic-ray driven dynamo (see Otmianowska-Mazur, this volume,
electronic edition). Turbulent dynamos can generate strong fields much
more quickly, but the dominant scale for field structure is then the
diffusive scale \citep{schekochihin04}, so there is no large-scale,
coherent field structure like that observed in galaxies. However, the
question of how the tangled field generated by a turbulent dynamo
evolves when embedded in a differentially rotating galaxy remains
unaddressed.

The simulations I have discussed in this paper do not answer this
question, but they do offer suggestive results that might point the
way towards the answer. Beginning with the simplest case, we note that
the in their stratified, shearing box MRI simulation with thermal
instability of \citet{piontek07} shows field growth to 3--4~$\mu$G in
the cold gas within 300~Myr, and then maintains that field strength
for another Gyr.  \citet{nishikori06} performed a 
three-dimensional, resistive MHD model of a global torus with
initially weak azimuthal field threading $10^5$~K gas. They found an
increase in magnetic energy of eight orders of magnitude over a period
of 200~Myr, and then maintain the field at that level for another
600~Gyr. 

\begin{figure*}
%Wang & Abel figure: field growth, configuration
\begin{center}
\includegraphics[width=0.65\textwidth]{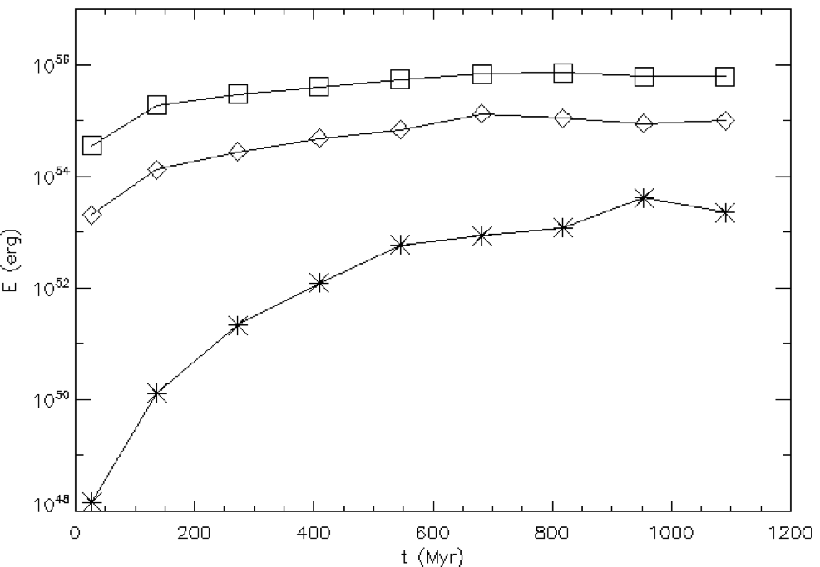}
\includegraphics[width=0.22\textwidth]{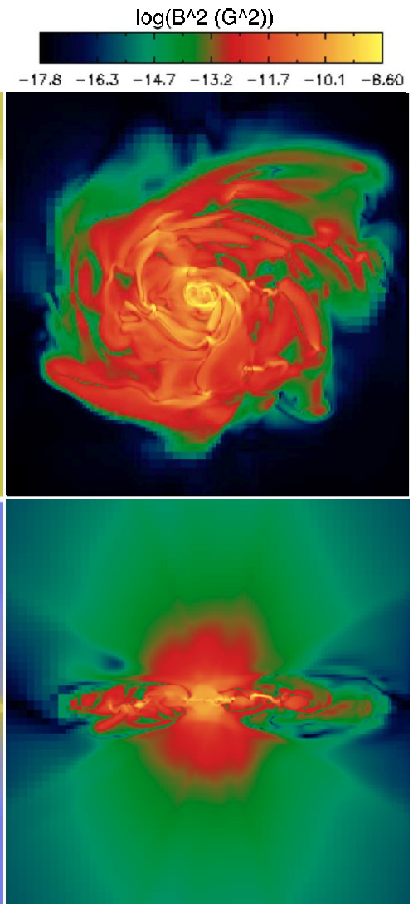}
\end{center}
\caption{\label{wang} {\em Left:} Time history from the AMR model of
\citet{wang08} of disk gas kinetic energy ({\em squares}), thermal
energy ({\em diamonds}), and magnetic energy ({\em asterisks}),
showing the strong growth of magnetic energy over time.  {\em Right:}
Horizontal and vertical slices of magnetic pressure after 1.088~Gyr of
evolution.  The plot square has a side of 11~kpc.  From
\citet{wang08}.}
\end{figure*}

\citet{wang08} find a similar behavior for the magnetic energy in
their simulation of disk formation in a collapsing halo, showing a
factor of $10^6$ amplification of field energy over 600~Myr, as shown
in Figure~\ref{wang}.  They
further demonstrate that the magnetic field in their disk indeed has
large-scale structure after 600~Myr, with several field reversals,
clear spiral structure, and field strengths in spiral arms exceeding
10--20~$\mu$G (Fig.~\ref{wang}).

\section{Summary}

The major points I have made in this paper are that the MRI may offer
a mechanism to maintain a minimum level of turbulence regardless of
the strength of star formation.  This might then support the picture
that gravitational instability acting on disks regulates the strength
of star formation in galaxies. The alternative view that magnetic
fields regulate star formation appears likely to be valid at least
morphologically, but it remains to be seen if the rate of star
formation requires magnetic regulation to explain.

A related question is how the large-scale magnetic field in galaxies
is generated.  Small-scale turbulent dynamos to generate field
strength, perhaps with large-scale coherent fields produced at the end
by classical dynamo action is an intriguing speculation that remains
consistent with the increasingly detailed models that I have reviewed
here.

\acknowledgments  I thank the organizers for their invitation to
speak.  This work was partly supported by NSF grant AST03-07854 and
NASA Origins of Solar Systems grant NNX07AI74G.


\begin{thebibliography}
%\setlength{\itemsep}{-\parsep}
%\setlength{\topsep}{-\parsep}
%\setlength{\partopsep}{-\parsep}

\bibitem[Balbus \& Hawley(1991)]{balbus91} Balbus, S. A., \& Hawley,
  J. F. 1991, ApJ, 376, 214
\bibitem[Balbus \& Hawley(1998)]{balbus98} Balbus, S. A., \& Hawley,
  J. F.. 1998, Rev.\ Mod.\ Phys., 70, 1
\bibitem[Dobbs \& Bonnell(2007)]{dobbs07} Dobbs, C. L., \& Bonnell,
  I. A. MNRAS, 374, 1115
\bibitem[Dziourkevitch et al.(2004)]{DER04} Dziourkevitch, N.,
  Elstner, D., \& R\"udiger, G. 2004, A\&A, 423, L29
\bibitem[Elmegreen(2002)]{elmegreen02} Elmegreen, B. G. 2002, ApJ,
  577, 206
\bibitem[Ferguson et al.(1998)]{ferguson98} Ferguson, A., Wyse,
  R. F. G., Gallagher, J. S., \& Hunter, D. A. 1998, ApJ, 506, L19 
\bibitem[Gammie(1992)]{gammie92} Gammie, C. F. 1992, PhD Thesis,
  Princeton University
\bibitem[Goldreich \& Lynden-Bell(1965)]{goldreich65} Goldreich, P.,
  \& Lynden-Bell, D. 1965, MNRAS, 130, 97
\bibitem[Hawley et al.(1995)]{hawley95} Hawley, J. F., Gammie, C. F.,
  \& Balbus, S. A. 1995, ApJ, 440, 742
\bibitem[Kennicutt(1998)]{kennicutt98} Kennicutt, R. C., Jr. 1998, ApJ, 498, 541
\bibitem[Kim et al.(2003)]{kim03} Kim, W.-T., Ostriker, E. C., \&
  Stone, J. M. 2003, ApJ, 599, 1157 
\bibitem[Kim \& Ostriker(2006)]{kim06} Kim, W.-T., \& Ostriker,
  E. C. 2006, ApJ, 646, 213
\bibitem[Kravtsov(2003)]{kravtsov03} Kravtsov, A. 2003, ApJ, 590, L1
\bibitem[Krumholz(2005)]{krumholz05} Krumholz, M. R., \& McKee, C. F. 2005,
  ApJ, 630, 250
\bibitem[Li et al.(2005a)]{lmk05} Li, Y., Mac Low, M.-M., \& Klessen,
  R. S. 2005, ApJ, 620, L19
\bibitem[Li et al.(2005b)]{lmk05b} Li, Y., Mac Low, M.-M., \& Klessen,
  R. S. 2005, ApJ, 626, 823
\bibitem[Li et al.(2006)]{lmk06}  Li, Y., Mac Low, M.-M., \& Klessen,
  R. S. 2006, ApJ, 639, 879
\bibitem[Mac Low(1999)]{ml99} Mac Low, M.-M. 1999, ApJ, 524, 169
\bibitem[Mac Low(2003)]{ml03} Mac Low, M.-M. 2003, in Turbulence and
  Magnetic Fields in Astrophysics, eds.\ E. Falgarone \& T. Passot
  (Heidelberg: Springer), 182
\bibitem[Martin \& Kennicutt(2001)]{mk01} Martin, C. L., \& Kennicutt,
  R. C., Jr. 2001, ApJ, 555, 301
\bibitem[Navarro, et al.(1996)]{navarro96} Navarro, J. F., Frenk,
  C. S., \& White, S. D. M. 1996, ApJ, 462, 563
\bibitem[Nishikori et al.(2006)]{nishikori06} Nishikori, H., Machida,
  M., \& Matsumoto, R. 2006, ApJ, 641, 862 
\bibitem[Petric \& Rupen(2007)]{petric07} Petric, A. O., \& Rupen,
  M. P. 2007, AJ, 134, 1952 
\bibitem[Piontek \& Ostriker(2004)]{piontek04} Piontek, R. A., \&
  Ostriker, E. C. 2004, ApJ, 601, 905
\bibitem[Piontek \& Ostriker(2005)]{piontek05} Piontek, R. A., \&
  Ostriker, E. C. 2005, ApJ, 629, 849
\bibitem[Piontek \& Ostriker(2007)]{piontek07} Piontek, R. A., \&
  Ostriker, E. C. 2007, ApJ, 663, 183
\bibitem[Rafikov(2001)]{rafikov01} Rafikov, R. R. 2001, MNRAS, 323,
  445
\bibitem[Rees(2006)]{rees06} Rees, M. J. 2006, Astron.\ Nach., 327,
  395
\bibitem[Schaye(2004)]{schaye04} Schaye, J. 2004, ApJ, 609, 667
\bibitem[Schekochihin et al.(2004)]{schekochihin04} Schekochihin,
  A. A., Cowley, S. C., Maron, J. L., \& McWilliams, J. C. 2004,
  Phys.\ Rev.\ Lett., 92, 084504
\bibitem[Sellwood \& Balbus(1999)]{sellwood99} Sellwood, J. A., \&
  Balbus, S. A. 1999, ApJ, 511, 660
\bibitem[Shetty \& Ostriker(2006)]{shetty06} Shetty, R., \& Ostriker,
  E. C. 2006, ApJ, 647, 997
\bibitem[Shu et al.(2007)]{shu07} Shu, F. H., Allen, R. J., Lizano,
  S., \& Galli, D. 2007, ApJ, 662, L75
\bibitem[Silk(1997)]{silk97} Silk, J. 1997, ApJ, 481, 703
\bibitem[Slyz et al.(2005)]{slyz05} Slyz, A. D., Devriendt, J. E. G.,
  Bryan, G., \& Silk, J. 2005, MNRAS, 356, 737
\bibitem[Toomre(1964)]{toomre64} Toomre, A. 1964, ApJ, 139, 1217
\bibitem[Wang \& Abel(2008)]{wang08}  Wang, P., \& Abel, T. 2008, ApJ,
  submitted (arXiv:0712.0872)
\bibitem[Yang et al.(2007)]{yang07} Yang, C.-C., Gruendl, R. A., Chu,
  Y.-H., Mac Low, M.-M., \& Fukui, Y. 2007, ApJ, 671, 374



\end{thebibliography}
\end{document}